\documentclass{aa}
\usepackage{rotating}
\usepackage{graphicx}
\usepackage{txfonts}
\begin{document}
\def\teff{$T_{\rm eff}$}
\def\logg{$\log g$}
\def\micro{$\xi$ }
\def\kms{km s$^{-1}$}
\def\p{$\pm$}
\def\vsini{$v\sin i$}

\title{The nature of the high Galactic latitude O-star HD\,93521: \newline new results from X-ray and optical spectroscopy\thanks{Based on observations collected at the Observatoire de Haute Provence (France) and with {\it XMM-Newton}, an ESA science mission with instruments and contributions directly funded by ESA member states and the USA (NASA).}}
\author{G.\ Rauw\thanks{Honorary Research Associate FRS-FNRS (Belgium)} \and T.\ Morel \and M.\ Palate}
\offprints{G.\ Rauw}
\mail{rauw@astro.ulg.ac.be}
\institute{Groupe d'Astrophysique des Hautes Energies, Institut d'Astrophysique et de G\'eophysique, Universit\'e de Li\`ege, All\'ee du 6 Ao\^ut, B\^at B5c, 4000 Li\`ege, Belgium}
\abstract{Owing to its unusual location and its isolation, the nature of the high Galactic latitude O9.5\,Vp object HD\,93521 is still uncertain.}{We have collected X-ray and optical observations to characterize the star and its surroundings.}{X-ray images and spectra are analyzed to search for traces of a recent star formation event around HD\,93521 and to search for the signature of a possible compact companion. Optical echelle spectra are analysed with plane-parallel model atmosphere codes, assuming either a spherical star or a gravity darkened rotationally flattened star, to infer the effective temperature and surface gravity, and to derive the He, C, N and O abundances of HD\,93521.}{The X-ray images reveal no traces of a population of young low-mass stars coeval with HD\,93521. The X-ray spectrum of HD\,93521 is consistent with a normal late O-type star although with subsolar metallicity. No trace of a compact companion is found in the X-ray data. In the optical spectrum, He and N are found to be overabundant, in line with the effect of rotational mixing in this very fast rotator, whilst C and O are subsolar. A critical comparison with the properties of subdwarf OB stars, indicates that, despite some apparent similarities, HD\,93521 does not belong to this category.}{Despite some ambiguities on the runaway status of the star, the most likely explanation is that HD\,93521 is a Population\,I massive O-type star that was ejected from the Galactic plane either through dynamical interactions or a result of a supernova event in a binary system.}
\keywords{Stars: early-type -- Stars: fundamental parameters -- Stars: massive -- Stars: individual: HD\,93521 -- X-rays: stars}
\authorrunning{Rauw et al.}
\titlerunning{HD\,93521: new results from X-ray and optical spectroscopy}
\maketitle
\section{Introduction \label{intro}}
O-type stars, the hottest and most massive main-sequence stars of Population I, are seldom found in isolation. They are preferentially found in open clusters and OB associations, which are therefore most likely to be their birth places (e.g.\ de Wit et al.\ \cite{deWit}). The few O-type stars that are not directly located inside a cluster or association are usually runaways and are hence believed to have been ejected from their parent cluster either by the kick of a supernova explosion in a binary system or as a result of dynamical interactions in a dense cluster core.

The high Galactic latitude O-type star HD\,93521 ($l_{\rm II} = 183.14^{\circ}$, $b_{\rm II} = 62.15^{\circ}$), located very far away from any known site of recent star formation, challenges this general picture. While the optical spectrum of HD\,93521 leads to an O9.5\,Vp classification, the nature of this star has been subject to debate over many years (see e.g.\ Ebbets \& Savage \cite{ES}, Irvine \cite{Irvine}, Howarth \& Smith \cite{HS}) and the case is still not settled.

HD\,93521 has one of the largest rotational velocities known among Galactic O-stars (390\,km\,s$^{-1}$, Rauw et al.\ \cite{Rauw}). The stellar wind has an apparently low terminal velocity and is likely heavily distorted into a Be-like decretion disk wind (Howarth \& Reid \cite{HR}, Bjorkman et al.\ \cite{Bjorkman}, Massa \cite{Massa}). In the optical domain, the wind produces emission features in the wings of the H$\alpha$ line, although they are far less prominent than in genuine Oe stars. In addition, HD\,93521 displays bi-periodic (1.75 and 2.89\,hr) absorption line profile variability that is commonly interpreted as the signature of two non-radial pulsation (NRP) modes with $l \approx 8 \pm 1$ and  $l \approx 4 \pm 1$(see Rauw et al.\ \cite{Rauw} and references therein). An alternative explanation for the optical line profile variations, that cannot totally be ruled out {\it a priori}, would be the effect of a compact companion (formed in the supernova explosion that ejected the system from the Galactic plane) orbiting the O-star and accreting material from its rotationally flattened wind. Indeed, in the case of Be/X-ray binaries, the decretion disk can be truncated by resonance with the orbit of the neutron star companion (Okazaki \& Negueruela \cite{ON}) and higher order resonance could possibly trigger periodic density waves in the disk that would contaminate the photospheric absoption lines via variable residual emission.

To help clarify the nature of HD\,93521, we have collected new optical and, for the first time, X-ray observations. Our observations are presented in Sect.\,\ref{obs}. Section\,\ref{cluster} is devoted to the study of the surroundings of HD\,93521 as seen in our X-ray images. The X-ray spectrum of HD\,93521 is analysed in Sect.\,\ref{EPIC93521}, whilst the optical spectra are studied in Sect.\,\ref{optspec} using plane-parallel atmosphere models assuming either spherical or rotationally flattened geometries for the star. The results are discussed in Sect.\,\ref{discussion} and Sect.\,\ref{conclusion} presents our conclusions.
 
\section{Observations \label{obs}}
\subsection{X-ray data}
A 40\,ksec X-ray observation was obtained on 2 - 3 November 2009 with the {\it XMM-Newton} satellite (Jansen et al.\ \cite{Jansen}). To reject optical and UV photons, the EPIC cameras (Turner et al.\ \cite{MOS}, Str\"uder et al.\ \cite{pn}) were used with the thick filter.

\begin{figure}[thb]
\begin{center}
\resizebox{9cm}{!}{\includegraphics{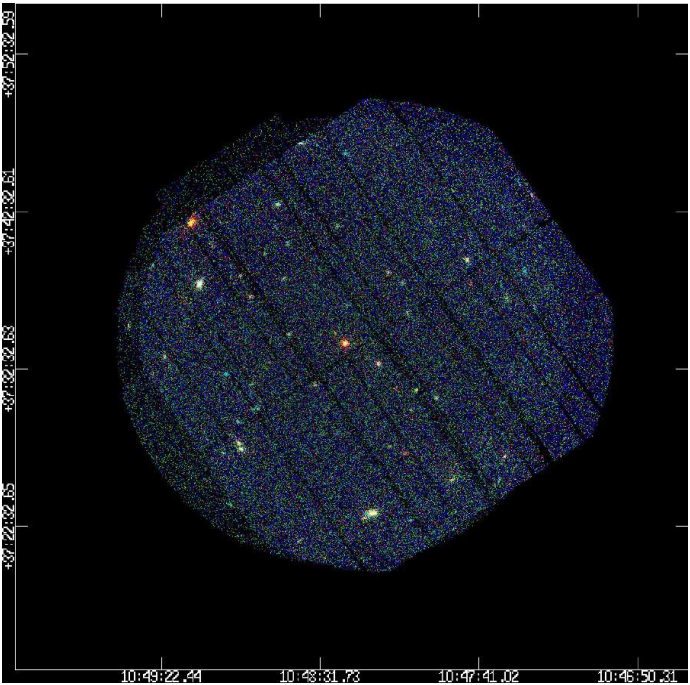}}
\end{center}
\caption{Energy-coded three-colour image built from our {\it XMM-Newton} observation of HD\,93521 (the reddish source near the center of the image). The red, green and blue colours correspond to the soft, medium and hard energy bands used throughout this paper (see text). The individual EPIC images were exposure corrected before they were combined. A colour version of the figure can be found in the electronic version of the journal. \label{3colours}}
\end{figure}

The raw data were processed with the SAS software version 10.0. The end of the observation was affected by the raise of a background flare. We discarded this part of the observation, ending up with an effective exposure time of 38\,ksec for EPIC-MOS1, 37\,ksec for EPIC-MOS2 and 30\,ksec for EPIC-pn. Images were built over soft (0.5 -- 1.0\,keV), medium (1.0 -- 2.0\,keV) and hard (2.0 -- 8.0\,keV) energy bands with a pixel size of $2.5$\arcsec $\times 2.5$\arcsec. These images were exposure corrected and combined into an energy-coded three-colour image of the field of view (Fig.\ \ref{3colours}). 

\subsection{Optical spectroscopy}
A series of twenty four optical spectra were obtained over four nights in April - May 2004 with the ELODIE \'echelle spectrograph (Baranne et al.\ \cite{Elodie}) at the 1.93\,m telescope at Observatoire de Haute Provence (OHP). ELODIE was an \'echelle spectrograph with a resolving power of $R \approx 42000$ over the 3850 -- 6800\,\AA\ wavelength domain. Our initial intention was to use these data for our study of the line profile variability (Rauw et al.\ \cite{Rauw}). However, due to the very poor weather conditions during this run, individual spectra lacked a sufficient signal-to-noise ratio and we thus decided to combine all the spectra into a single high-quality spectrum. This average spectrum, with a signal to noise ratio (S/N) of 390 at 5500\,\AA, was obtained from data collected at many different phases of the line profile variation cycles and should hence be relatively free of any signatures of the latter.  

Another \'echelle spectrum was obtained in November 2010 with the SOPHIE spectrograph (Perruchot et al.\ \cite{Sophie}) also at the 1.93\,m telescope at OHP. SOPHIE covers the wavelength domain from 3870 to 6940\,\AA. The spectrum was reduced with IRAF. This spectrum is of higher resolution than the ELODIE ones with a resolving power, $R \approx 74\,000$, as measured from calibration lamps. The S/N of this spectrum is $\approx 310$ at 5500\,\AA.

\subsection{H$\alpha$ imaging \label{Halpha}}
Since there are apparently no H$\alpha$ images available for the region around HD\,93521, we collected some images with the help of two amateur astronomers, Gaston Dessy and Jo\"el Bavais. For this purpose, G.\ Dessy used a TMB-92 9.2\,cm refractor with a focal length of 40.8\,cm equipped with an Atik 16IC CCD ($659 \times 494$\,pixels), and for a second run with an Atik 4000M CCD ($2048 \times 2048$ pixels). Both CCDs have pixel sizes of $7.4 \times 7.4$\,$\mu$m$^2$, corresponding to $3.7$\arcsec $\times 3.7$\arcsec on the sky. J.\ Bavais used a SkyWatcher 80/600 ED refractor equipped with an Atik 314L CCD ($1392 \times 1040$\,pixels of $6.45 \times 6.45$\,$\mu$m$^2$ corresponding to $2.2$\arcsec $\times 2.2$\arcsec on the sky). In all cases, commercial H$\alpha$ (70\,\AA\ bandwidth) and $V$-band filters manufactured by Baader Planetarium were used. For each set-up, integration times were 1\,hour in H$\alpha$ (split into 12 exposures of 5\,min each) and 15\,min in $V$ (split into 5 exposures of 3\,min each, except for the last run where no $V$-band images were taken). All data were processed with the MIDAS software developed at ESO. The surface brightness of the sky measured on our images amounts to about 17.8\,mag\,arcsec$^{-2}$ in $V$. After subtracting a flat sky background, the residual 1-$\sigma$ fluctuations in the $V$ band images correspond to a surface brightness around 22.0\,mag\,arcsec$^{-2}$. The H$\alpha$ images are about 1.5\,mag shallower than the $V$-band data. No trace of a nebular emission was found on the images before or after sky subtraction.

\section{The surroundings of HD\,93521 \label{cluster}}
The position of HD\,93521 far away from the Galactic plane and from any open cluster naturally raises the question of the origin of this star. There are currently only a few known examples of massive stars that have likely formed in isolation. For instance, de Wit et al.\ (\cite{deWit}) conclude that $4 \pm 2$\% of the Galactic O-stars could have formed outside clusters\footnote{In the same context, Bressert et al.\ (\cite{Bressert}) recently identified a sample of massive stars, earlier than spectral type O7, within a projected distance of 125\,pc around the 30\,Dor complex, that are not associated with any cluster and are thus candidates for massive stars formed in isolation. It has to be stressed though that the case of HD\,93521 requires a far more extreme form of isolation than considered by Bressert et al.\ (\cite{Bressert}).}.

Models for massive star formation based on the competitive accretion scenario imply that a high-mass star must be associated with a population of low-mass stars (Bonnell et al.\ \cite{Bonnell})\footnote{Although dynamical interaction could lead the sparsest clusters to disperse rather quickly, on time scales of a few Myr.}. Core accretion models of massive stars on the contrary allow rather isolated O-stars to form (Krumholz et al.\ \cite{Krumholz}).
Isolated O-type stars that are still in their formation region are surrounded by residual gas that can be seen through H$\alpha$ observations (Lamb et al.\ \cite{Lamb}, Selier et al.\ \cite{Selier}) and some of them are found in sparse clusters with less than ten lower-mass companions (Lamb et al.\ \cite{Lamb}). Monte Carlo simulations by Lamb et al.\ (\cite{Lamb}) indicate that the existence of such sparse clusters is more in favour of the core accretion models and suggest that `clusters are built stochastically by randomly sampling stars from a universal initial mass function (IMF)'.
 
\begin{table*}[htb]
\caption{Spectral properties of the brightest X-ray sources other than HD\,93521.\label{fitsX}}
\begin{tabular}{l c c c c c c c c}
\hline
Source & Model & $N_{\rm H}$ & kT & $\Gamma$ & $f_X$ (0.5 - 10\,keV) & $f_X^{\rm unabs}$ (0.5 - 10\,keV) & $\chi^2_{\nu}$ & d.o.f.\\
       &     & ($10^{22}$\,cm$^{-2}$) & (keV) &    & (erg\,cm$^{-2}$\,s$^{-1}$) & (erg\,cm$^{-2}$\,s$^{-1}$) &  & \\
\hline
\vspace*{-2mm}\\
BD$+38^{\circ}$\,2183 & {\tt wabs*apec} & $< 0.020$ & $0.56^{+.05}_{-.07}$ & -- & $1.11\times 10^{-13}$ & $1.11\times 10^{-13}$ & 0.90 & 11 \\
\vspace*{-2mm}\\
BZQ\,J1049+3737 & {\tt wabs*power} & $0.024^{+.020}_{-.016}$ & -- & $1.67^{+.10}_{-.11}$ & $2.12\times 10^{-13}$ & $2.18\times 10^{-13}$ & 1.05 & 65 \\
\vspace*{-2mm}\\
source 4 & {\tt wabs*power} & $0.045^{+.018}_{-.015}$ & -- & $2.02^{+.13}_{-.12}$ & $2.27\times 10^{-13}$ & $2.43\times 10^{-13}$ & 1.54 & 73\\
\vspace*{-2mm}\\
\hline
\end{tabular}
\end{table*}
\begin{figure}[thb]
\begin{center}
\resizebox{9cm}{!}{\includegraphics[angle=-90]{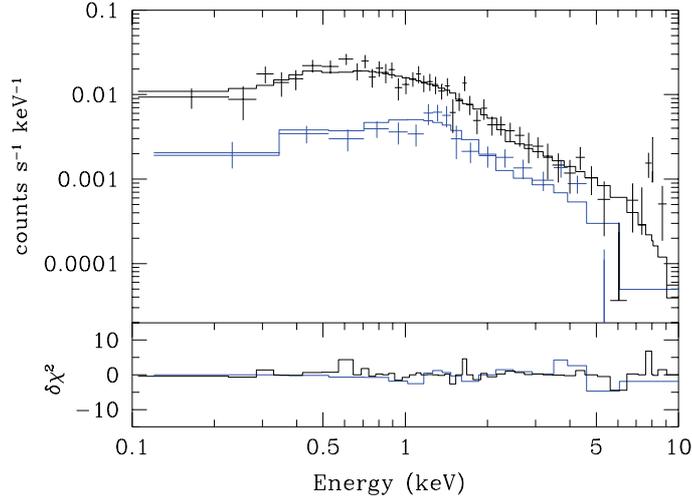}}
\end{center}
\caption{Best-fit absorbed power-law model (see Table\,\ref{fitsX}) of the X-ray spectrum of the blazar BZQ\,J1049+3737. The black and blue data points refer to the pn and MOS2 data, respectively. The lower panel shows the contribution of the various energy bins to the $\chi^2$ of the fit times the sign of the difference between the data and the model.\label{qso}}
\end{figure}

X-ray observations can help us uncover low-mass, optically dim members of very young open clusters. Indeed, X-ray images of massive stars in dense open clusters usually reveal a rather rich population of X-ray bright, low-mass pre-main sequence stars (e.g.\ Damiani et al.\ \cite{Damiani}, Sana et al.\ \cite{ngc6231}). This statement also holds to some extent for rather scarce open clusters such as NGC\,6383 (Rauw et al.\ \cite{ngc6383}) and to less densely populated OB associations (e.g.\ Rauw \cite{cygob2}). 
The brightness of very young low-mass stars in the X-ray domain makes this the ideal energy range to distinguish between foreground or background field stars and genuine cluster members coeval with the massive stars. This technique was successfully applied to reveal the population of low-mass pre-main sequence (PMS) stars associated with two rather isolated B0 stars in the diffuse H\,{\sc ii} regions S\,255 and S\,257 (Mucciarelli et al.\ \cite{MPZ}).

By analogy with the two B0 stars investigated by Mucciarelli et al. (\cite{MPZ}), one expects about 300 low-mass pre-main sequence stars associated with HD\,93521 if this stellar population follows a standard IMF\footnote{This estimate is in line with the relation between the total cluster mass and the mass of its most massive member as proposed by Weidner et al.\ (\cite{Weidner}). Indeed, applying this relation to HD\,93521, one would expect a cluster with a total mass in the range 100 -- 150\,M$_{\odot}$.}. We can now estimate what fraction of these low-mass objects would be detected with our X-ray observation and compare this to the actual number of sources in the field of view.

The three-colour image of the field of view around HD\,93521 (Fig.\,\ref{3colours}) does not reveal any obvious concentration of X-ray sources around HD\,93521. Applying the SAS source detection routines with a significance threshold\footnote{This implies a probability of $\leq e^{-10}$ that a random Poissonian fluctuation could have caused the observed source counts inside the $5 \times 5$\,pixel$^2$ detection cell. Over the entire EPIC field of view, the expected number of spurious detections amounts to about 2\% of the total number of detected sources.} of 10 yields a total of 55 sources. After visual inspection, 10 sources were removed from the list because their detection was affected by gaps between CCDs or because they were at the edge of the field of view. Two more faint sources that were apparently missed by the detection algorithm were added to our final list of 47 X-ray sources. Assuming an optically thin thermal plasma model with a typical temperature of 1\,keV, our detection limit corresponds to an observed flux of about $3 \times 10^{-15}$\,erg\,cm$^{-2}$\,s$^{-1}$. At an adopted distance of HD\,93521 of 1.2\,kpc (see below), this flux would correspond to an X-ray luminosity of $5 \times 10^{29}$\,erg\,s$^{-1}$. If we compare this number with the cumulative X-ray luminosity functions of the Orion Nebular Cluster (Preibisch \& Feigelson \cite{COUP}), we estimate that we should detect about 90, 85 and 50\% of the PMS stars in the mass ranges 0.9 -- 1.2, 0.5 -- 0.9 and 0.1 -- 0.5\,M$_{\odot}$ respectively. Therefore, we would expect to observe of order 100 - 200 X-ray sources in the vicinity of HD\,93521. This is clearly not the case here. 
 
To assess the impact of background X-ray sources, we have repeated the detection chain on two different energy bands 0.5 -- 2.0\,keV and 2.0 -- 8.0\,keV. In the 0.5 -- 2.0\,keV band, 42 sources are detected with a limiting flux of order 3 -- $4 \times 10^{-15}$\,erg\,cm$^{-2}$\,s$^{-1}$. From the $\log{N}$ -- $\log{S}$ relation of Giacconi et al.\ (\cite{Giacconi}) derived from the {\it Chandra} Deep Field South, we would expect about 50 detections of extragalactic point sources in this energy band and with this flux limit. This result implies that the bulk of the detected sources must be extragalactic background objects, unrelated to HD\,93521. Very similar conclusions are reached with the results for the 2.0 -- 8.0\,keV band. This conclusion is further supported by the fact that only five X-ray sources in our field of view have a 2MASS near-IR counterpart within a correlation radius of 4\,arcsec. Out of these five sources, four are among the brightest X-ray sources: HD\,93521 itself, the blazar [MGL2009]BZQ\,J1049+3737 (Massaro et al.\ \cite{Massaro}), the star BD$+38^{\circ}$\,2183 (with a SIMBAD spectral type F8), and an unknown source at 10:48:14.9 $+37$:23:24 (hereafter source 4).
 
We have analysed the EPIC spectra of the three brightest sources in addition to HD\,93521. The results are listed in Table\,\ref{fitsX}. The blazar and the unknown source are best fitted with a power-law model (see Fig.\,\ref{qso}), suggesting that the latter is most probably also related to an AGN. The spectrum of the late-type star is well represented by a single temperature optically thin thermal plasma model.

We thus conclude that our {\it XMM-Newton} data do not reveal any evidence for a lower mass stellar population that could be associated with the formation of the O-type star HD\,93521. This is in line with the lack of nebular emission from residual gas in our H$\alpha$ images (see Sect.\,\ref{Halpha}).

\section{The EPIC spectrum of HD\,93521 \label{EPIC93521}}
HD\,93521 is detected as a moderate X-ray source with net count rates of $7.3 \times 10^{-3}$, $8.2 \times 10^{-3}$ and $3.6 \times 10^{-2}$ cts\,s$^{-1}$ for EPIC-MOS1, MOS2 and pn, respectively.  
The EPIC spectra of HD\,93521 were analysed using the {\tt xspec} software (version 12.6.0, Arnaud \cite{Arnaud}). Unless stated otherwise, the interstellar neutral hydrogen (H\,{\sc i} + H$_2$) column density was set to $1.3 \times 10^{20}$\,cm$^{-2}$ (Bohlin et al.\ \cite{BSD}). To account for the possible presence of an additional circumstellar absorption column due to the partially ionized stellar wind, we further included an ionized wind absorption model (Naz\'e et al.\ \cite{wind}) with the wind column density treated as a free parameter.

We have tested a variety of models that are potentially adequate depending on the actual nature of HD\,93521.
 
Medium resolution X-ray spectra of normal O-type stars are usually well fitted with absorbed optically thin thermal plasma models (see e.g.\ Naz\'e \cite{YN}). A model with a single plasma component does not fit the EPIC spectra of HD\,93521 simultaneously at low and high energies. We have therefore tested two-temperature apec models (Smith et al.\ \cite{apec}), either with solar metallicity or with the global metallicity of the plasma taken as a free parameter (see models $[1]$ and $[2]$ in Table\,\ref{fitsHD93521}). In the latter case, the best fit is achieved with a metallicity near 0.22. This fit provides a significant improvement over the solar metallicity solution. 

\begin{figure}[thb]
\begin{center}
\resizebox{9cm}{!}{\includegraphics[angle=-90]{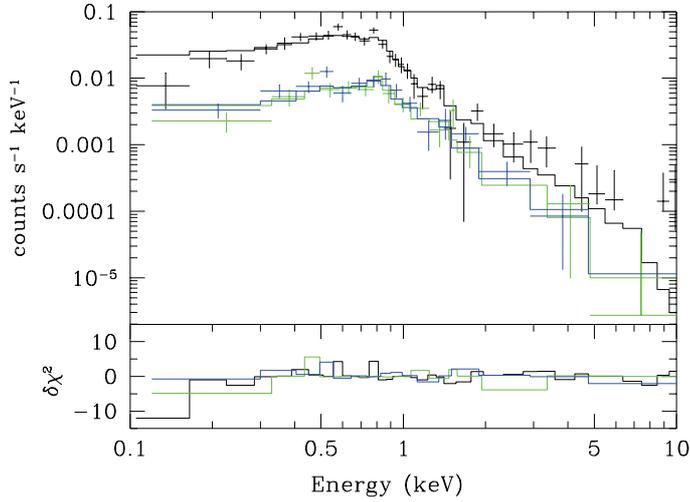}}
\end{center}
\caption{Same as Fig.\,\ref{qso}, but for the best-fit model $[3]$ (see Table\,\ref{fitsHD93521}) of the X-ray spectrum of HD\,93521. The black, green and blue data points refer to the pn, MOS1 and MOS2 data respectively.\label{modelHD93521}}
\end{figure}

Using the He and CNO abundances derived from our best fit of the optical spectrum (see Sect.\,\ref{optspec}), we have then tested a two temperature apec model with variable abundances (vapec). The He, C, N and O abundances were fixed at respectively 2.14, 0.19, 1.51 and 0.30 times solar. As a first step, all other elements were taken to have solar abundances. This led to a relatively poor fit with $\chi^2_{\nu} = 1.52$. Apart from C, N and O, Fe and Ne have strong lines in the X-ray domain covered by our EPIC data. Therefore, we allowed the abundances of these two elements to vary in the fit. This results in an improved fit with Ne and Fe abundances of respectively $0.28^{+.20}_{-.15}$ and $0.43^{+.14}_{.15}$ relative to solar (see Fig.\,\ref{modelHD93521} and model $[3]$ in Table\,\ref{fitsHD93521}). Assuming that Ne and Fe have the same abundances as O (i.e.\ 0.30 solar), which is our best indicator of the metallicity as the O abundance is only little affected by the CNO process, yields an equivalent fit (see model $[4]$ in Table\,\ref{fitsHD93521}).  

Whatever the metallicity, we find that the spectrum can be represented by two plasma components of temperatures near 0.3 and 3.0\,keV, though the hotter component is poorly constrained (especially towards higher temperatures). We further find that a significant wind absorption is needed to provide a reasonable fit and that the interstellar absorption corrected flux corresponding to thermal plasma models is of order $6.0$ -- $6.5 \times 10^{-14}$\,erg\,cm$^{-2}$\,s$^{-1}$. Assuming a distance of 1.2\,kpc, this flux corresponds to an X-ray luminosity of 1.0 -- 1.1 $\times 10^{31}$\,erg\,s$^{-1}$, which translates into an $L_{\rm X}/L_{\rm bol}$ ratio of 8.7 -- 9.4 $\times 10^{-8}$. This number is towards the lower end of, but compatible with, the range of $L_{\rm X}/L_{\rm bol}$ values measured for normal O-type stars with {\it XMM-Newton} (Naz\'e \cite{YN}).

A plasma component as hot as 3\,keV is somewhat surprising for intrinsic X-ray emission by a single ordinary O-type star. This is especially true if the star has indeed a low equatorial wind velocity, of order $v_{\infty} = 400$\,km\,s$^{-1}$, as derived by Howarth et al.\ (\cite{Howarth}, although the polar wind component is likely much faster, $v_{\infty} = 2000$\,km\,s$^{-1}$, see Howarth \& Reid \cite{HR}). Naz\'e (\cite{YN}) found that many O-type stars feature a faint hot plasma contribution around 2\,keV, but only a few objects have a plasma that is actually as hot as in the case of HD\,93521. Over recent years, a new category of intrinsic hard X-ray sources associated with early-type stars was identified, the so-called $\gamma$\,Cas analogs (e.g.\ Smith et al.\ \cite{Smith}, Rakowski et al.\ \cite{Rakowski}). These are late O or early B-type emission-line stars which have X-ray spectra dominated by plasma components with kT up to 12\,keV, located rather close to the Be star and its disk. However, unlike HD\,93521, these objects are highly overluminous (by a factor $\approx 40$, Rakowski et al.\ \cite{Rakowski}) with respect to the typical $L_{\rm X}/L_{\rm bol}$ ratio of OB stars. Therefore, HD\,93521 has rather normal X-ray properties for an O-type star, except for the somewhat higher secondary plasma temperature.

As an alternative, we have tested the possibility that the hard emission might be non-thermal. Such a non-thermal emission could arise either from an accretion process in a binary system featuring a compact companion or from inverse Compton scattering of photospheric UV photons, if the wind of HD\,93521 contains a population of relativistic electrons. The fit is of similar quality to the best fit purely thermal model (see model $[5]$ in Table\,\ref{fitsHD93521}). The softer emission still requires a thermal plasma component at about 0.3\,keV. For this kind of model, we have also attempted to let the metallicity of the apec component vary during the fit. There is no improvement of the fit quality and, apart from the metallicity which goes to $0.29^{+.77}_{-.14}$\,Z$_{\odot}$ and the normalization parameter, there are no changes in the best fit parameters. Finally, we also tested a model with the individual abundances of He, C, N and O set to the values derived from our best fit of the optical spectrum and letting the Ne and Fe abundances vary. Again, there is no improvement of the fit and the abundances of Ne and Fe converge to values very similar to those obtained with model $[3]$.

\begin{table*}[htb]
\caption{Spectral fits of HD\,93521.\label{fitsHD93521}}
\begin{tabular}{c c c c c c c c c c c c}
\hline
Model & $\log{N_{\rm wind}}$ & kT$_1$ & norm$_1$ & kT$_2$ & $\Gamma_2$ & norm$_2$ & Z & $f_X$  & $f_X^{\rm unabs}$ & $\chi^2_{\nu}$ (d.o.f.)\\
      & (cm$^{-2}$) & (keV) & & (keV) &  & & Z$_{\odot}$ &\multicolumn{2}{c}{($10^{-14}$\,erg\,cm$^{-2}$\,s$^{-1}$)} & \\
\hline
\vspace*{-2mm}\\
$[1]$ & $21.63^{+.08}_{-.09}$ & $0.27^{+.01}_{-.01}$ & $(0.89^{+0.14}_{-0.14})\,10^{-4}$ & $3.01^{+3.51}_{-0.96}$ & -- & $(0.18^{+.05}_{-0.05})\,10^{-4}$ & 1 (fixed) & $6.20$ & $6.52$ & 1.49 (65) \\
\vspace*{-2mm}\\
$[2]$ & $21.54^{+.12}_{-.21}$ & $0.28^{+.02}_{-.01}$ & $(2.65^{+1.44}_{-0.93})\,10^{-4}$ & $2.94^{+7.32}_{-1.14}$ & -- & $(0.22^{+0.09}_{-0.09})\,10^{-4}$ & $0.22^{+.12}_{-.09}$ &  $5.89$ & $6.19$ & 1.34 (64) \\
\vspace*{-2mm}\\
$[3]$ & $21.37^{+.18}_{-.23}$ & $0.29^{+.03}_{-.02}$ & $(1.15^{+0.27}_{-0.29})\,10^{-4}$ & $2.96^{+10.05}_{-1.35}$ &  -- & $(0.14^{+.05}_{-0.06})\,10^{-4}$ & model & $5.57$ & $6.02$ & 1.32 (63) \\
\vspace*{-2mm}\\
$[4]$ & $21.26^{+.17}_{-.33}$ & $0.30^{+.01}_{-.01}$ & $(1.08^{+0.35}_{-0.26})\,10^{-4}$ & $4.03^{+63.7}_{-2.30}$ & -- & $(0.12^{+0.05}_{-0.04})\,10^{-4}$ & model &  $5.72$ & $6.00$ & 1.31 (65) \\
\vspace*{-2mm}\\
$[5]$ & $21.54^{+.09}_{-.11}$ & $0.28^{+.01}_{-.01}$ & $(0.62^{+0.23}_{-0.17})\,10^{-4}$ & -- & $2.54^{+.29}_{-.36}$ & $(0.10^{+.03}_{-.03})\,10^{-4}$ & 1 (fixed) & $5.97$ & $6.27$ & 1.35 (65) \\
\vspace*{-2mm}\\
$[6]$ & -- & $0.152^{+.008}_{-.006}$ & $(8.5^{+0.8}_{-0.7})\,10^{-7}$ & -- & $1.44^{+.36}_{-.29}$ & $(3.0^{+1.8}_{-1.5})\,10^{-6}$ & -- & $6.15$ & $6.41$ & 1.46 (66) \\
\vspace*{-2mm}\\
\hline
\end{tabular}
\tablefoot{All fits were performed with the interstellar neutral hydrogen column density set to $1.3 \times 10^{20}$\,cm$^{-2}$. Models $[1]$ and $[2]$ correspond to {\tt wabs*wind*apec(2T)} with solar and free metallicity respectively. Models $[3]$ and $[4]$ correspond to {\tt wabs*wind*vapec(2T)} with the abundances of He, C, N and O set to 2.14, 0.19, 1.51 and 0.30 times solar. In model $[3]$, the abundances of Ne and Fe are free (see text), whilst they are fixed at 0.30 solar in model $[4]$. Models $[5]$ and $[6]$ correspond to {\tt wabs*wind*(apec+power)} and {\tt wabs*(bbody+power)} respectively. The fluxes are evaluated over the energy range 0.5 -- 10\,keV. The normalization of apec models are defined as $\frac{10^{-14}}{4\,\pi\,d^2}\,\int n_e\,n_H\,dV$ where $d$ is the distance in cm, $n_e$ and $n_H$ are the electron and proton densities and $V$ is the volume of the emitting plasma. The normalization parameters of the power law and black-body component are respectively the number of photons keV$^{-1}$\,cm$^{-2}$\,s$^{-1}$ at 1\,keV and the source luminosity in units $10^{39}$\,erg\,s$^{-1}$ divided by the square of the distance in units of 10\,kpc.}
\end{table*}

Finally, we have tested a model consisting of a black-body and a power law. This kind of model was found to adequatly describe the EPIC spectra of the sdO + compact companion system HD\,49798 (Tiengo et al.\ \cite{Tiengo}). The results are listed as model $[6]$ in Table\,\ref{fitsHD93521}. If we keep the interstellar column fixed, the best fit is of lower quality than what can be achieved with thermal plasma models. We stress that the best-fit black-body component of HD\,93521 would be much hotter (kT = 0.15 vs.\ 0.034\,keV) and the power law significantly steeper ($\Gamma = 1.4$ vs.\ 2.0) than in the case of HD\,49798 (Tiengo et al.\ \cite{Tiengo}). A significantly better fit ($\chi^2_{\nu} = 1.08$) could be obtained with the black-body + power law model if the interstellar column density would be treated as a free parameter. However, in this case, the fitted ISM column density would reach $0.14 \times 10^{22}$\,cm$^{-2}$, i.e.\ it would exceed the observationally determined column density by a factor 10. Such a large discrepancy seems rather unlikely.

\begin{figure}[thb]
\begin{center}
\resizebox{8cm}{!}{\includegraphics{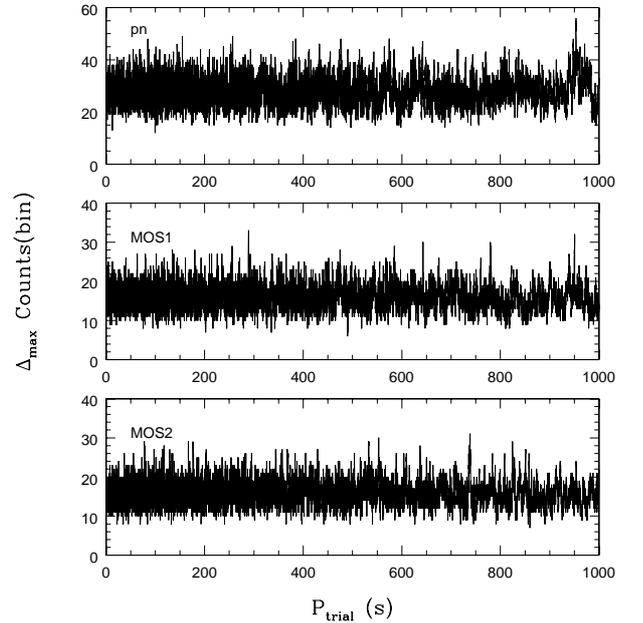}}
\end{center}
\caption{Maximum difference between the number of X-ray counts recorded per 0.05 phase bin as a function of the trial period. The various panels yield the results for the three EPIC detectors. The total number of counts (source + background) recorded in the source region were 363, 362 and 1143 respectively for the MOS1, MOS2 and pn detectors.\label{tempofast}}
\end{figure}

To further test the possibility that HD\,93521 could be a binary system hosting a compact companion (either a white dwarf or a neutron star), we have searched the X-ray data for periodicities that could be related to X-ray pulses from the compact companion accreting material from the O9.5 component. For this purpose, we have extracted event lists from the source region of HD\,93521 (containing photons from the source and the background\footnote{The source accounts for roughly 70\% of the photons, the remainder are due to the background.}) for each of the three EPIC instruments. The arrival times of these photons were corrected into the barycentric frame of reference. We then folded the arrival times of all the photons with a trial period and built a histogram of the resulting arrival phases (adopting phase bins of 0.05). We finally compute the maximum difference between the number of counts in the phase bins of the histogram. The whole process was repeated for 10000 trial periods between 1.0 and 1000\,s\footnote{In the full frame modes used in our observation, the pn and MOS detectors have a time resolution of 0.0734 and 2.6\,s respectively.} and the maximum difference in the number of counts was plotted as a function of the trial period. If there was a periodic signal in the X-ray data of HD\,93521, we would expect a peak in this diagram. No such peak is found in our data (see Fig.\,\ref{tempofast}) and we thus conclude that the X-ray emission of HD\,93521 does not contain any signature of short-period pulsations that could hint at the presence of an accreting compact companion.  

\section{Analysis of the optical spectrum\label{optspec}}
\subsection{Previous work}
The optical spectrum of HD\,93521 has been investigated previously with model atmosphere codes. The most sophisticated approach so far, was the work of Howarth \& Smith (\cite{HS}). These authors used non-LTE, hydrostatic, plane-parallel H/He model atmospheres to analyse the spectrum of HD\,93521, accounting for the variation of the local gravity (and hence temperature) as a function of stellar latitude at the surface of a rotationally distorted star. Howarth \& Smith (\cite{HS}) assumed an inclination of $90^{\circ}$ and their best-fit model yielded $T_{\rm eff}^p = 38000 \pm 1500$\,K,  $T_{\rm eff}^e = 33500$\,K, $\log{g}^p = 3.9 \pm 0.1$ and $\log{g}^e = 3.5$ (in cgs units), a helium abundance by number of $y = 0.18 \pm 0.03$ and a star rotating at 90\% of its critical velocity. Here the `$p$' and `$e$' superscripts refer to the parameters respectively at the poles and at the equator of the star.

\begin{table*}[thb]
\centering
\caption{Atmospheric parameters and metal abundances (on the scale in which $\log \epsilon$[H]=12) for the reference stars and for HD\,93521.}
\begin{tabular}{l|cc|cccccccc}
\hline\hline
                                & \teff \ [K]   & \logg      & $y$       & $\log \epsilon$(C)  & $\log \epsilon$(N)  & \multicolumn{2}{c}{$\log \epsilon$(O)} & [N/C]       & [N/O]      \\
                                &               &            &            & [4060-4082 \AA]     & [4995-5011 \AA]     & [4060-4082 \AA]    & [4691-4709 \AA]   &             &            \\\hline
\bf{10\,Lac }                    & \bf{34\,000}  & \bf{4.15}  & \bf{0.097} & \bf{8.31}           & \bf{7.55}           & \bf{8.39}          & \bf{8.39}         & \bf{--0.76} & \bf{--0.84}\\
                                & 34\,254       & 4.20       & 0.079      & 8.22                & 7.44                & 8.34               & 8.28              & --0.78      & --0.87     \\
\vsini \ = 390 \kms             & 33\,975       & 4.27       & 0.075      & 8.30                & 7.52                & 8.42               & 8.34              & --0.78      & --0.86     \\\hline
\bf{$\tau$\,Sco}                 & \bf{31\,500}  & \bf{4.05}  & \bf{0.085} & \bf{8.19}           & \bf{8.15}           & \bf{8.62}          & \bf{8.62}         & \bf{--0.04} & \bf{--0.47}\\
                                & 31\,088       & 4.15       & 0.101      & 8.18                & 7.92                & 8.48               & 8.48              & --0.26      & --0.56     \\
\vsini \ = 390 \kms             & 30\,983       & 4.18       & 0.096      & 8.42                & 8.08                & 8.74               & 8.64              & --0.34      & --0.61     \\\hline 
\bf{HD\,57682}                   & \bf{33\,000}  & \bf{4.00}  & \bf{0.096} & \bf{8.20}           & \bf{7.52}           & \bf{8.31}          & \bf{8.31}         & \bf{--0.68} & \bf{--0.79}\\
                                & 33\,958       & 4.17       & 0.096      & 8.14                & 7.58                & 8.34               & 8.36              & --0.56      & --0.77     \\
\vsini \ = 390 \kms             & 31\,814       & 3.97       & 0.079      & 8.04                & 7.52                & 8.54               & 8.38              & --0.52      & --0.94     \\\hline\hline 
\bf{HD\,93521, sph.\ symm.}     &               &            &            &                     &                     &                    &                   &             &            \\
SOPHIE                          & 30\,944       & 3.72       & 0.182      & 7.70                & 8.00                & 8.20               & 8.16              & +0.30       & --0.18     \\ 
ELODIE                          & 30\,892       & 3.62       & 0.174      & 7.42                & 7.94                & 8.28               & 8.32              & +0.52       & --0.36     \\
Mean                            & 30\,918       & 3.67       & 0.178      & 7.56                & 7.97                & 8.24               & 8.24              & +0.41       & --0.27     \\\hline
\bf{HD\,93521, non sph.} &               &            &            &                     &                     &                    &                   &             &            \\
SOPHIE                          &               &            &            & 7.74                & 8.03                & 8.22               & 8.22              & +0.29       & --0.19     \\ 
ELODIE                          &               &            &            & 7.49                & 7.97                & 8.32               & 8.34              & +0.48       & --0.36     \\
Mean                            &               &            &            & 7.62                & 8.00                & 8.27               & 8.28              & +0.38       & --0.27     \\\hline
\end{tabular}
\label{tab_abundances}  
\tablefoot{The reference values based on classical methods are shown in boldface. For the standard stars, the second line yields the results of our analysis of the actual spectra of these stars using the same approach as for HD\,93521, whilst the third line gives the corresponding results for the spectra degraded to the \vsini \ of HD\,93521. The solar [N/C] and [N/O] abundance ratios are --0.60 and --0.86, respectively (Asplund et al. \cite{asplund09}).}
\end{table*}
\begin{figure}[h!]
\centering
\includegraphics[width=90mm]{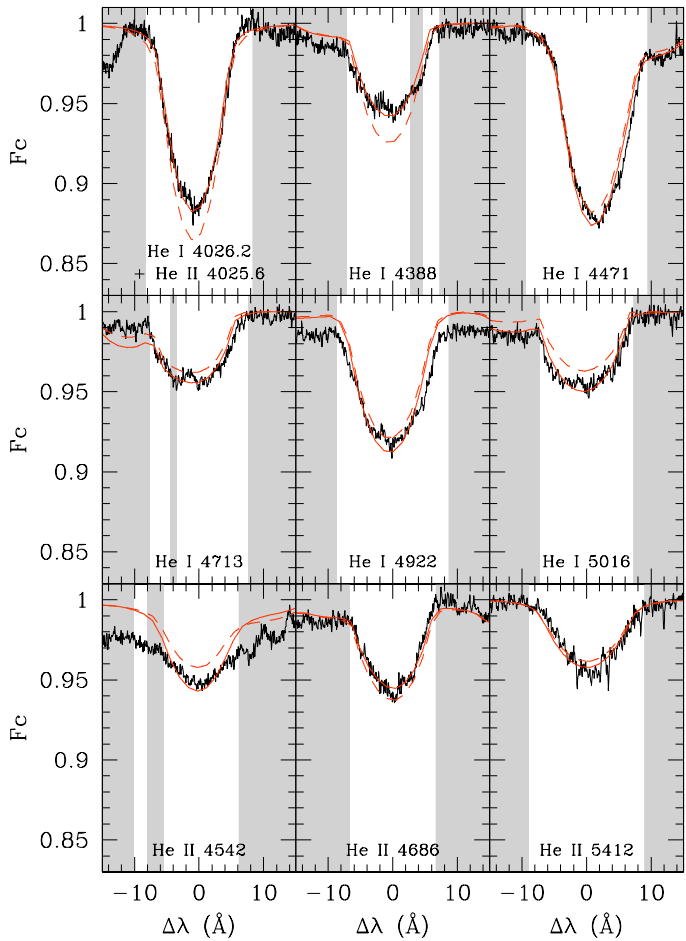}
\caption{Comparison for HD 93521 between the SOPHIE ({\it black}) and
best-fitting He synthetic line profiles ({\it solid red}). The dashed,
red lines show the line profiles computed for the final parameters
derived from the analysis of the SOPHIE spectrum (Table
\ref{tab_abundances}). The light grey-shaded areas delineate the regions
where the quality of the fit has been evaluated.}
\label{fig_he}
\end{figure}
\begin{figure*}
\centering
\includegraphics[width=140mm]{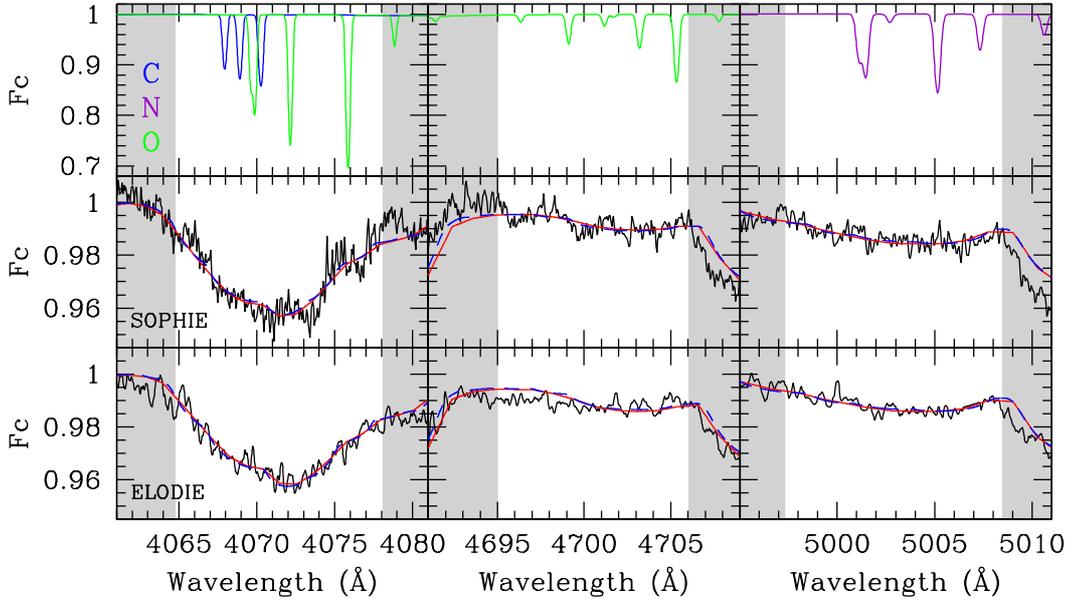}
\caption{Comparison for HD\,93521 between the observed ({\it black}) and best-fitting synthetic metal line profiles under the assumption of a spherical star ({\it red}) and a gravity darkened rotationally flattened star ({\it blue}). The light grey-shaded areas delineate the regions where the quality of the fit has been evaluated. The top panels show the synthetic profiles computed for \teff \ = 31\,000 K, \logg \ = 3.7 dex, \micro \ = 10 \kms \ and the abundances obtained using the ELODIE spectrum and assuming a spherical star.}
\label{fig_fits}  
\end{figure*}

\subsection{1-D model atmospheres}\label{sect_1D}
In this section we first analyse the spectrum assuming a spherically symmetric star to derive stellar parameters as well as abundances of He, C, N and O. These results are used in Sect.\,\ref{sect_2D} as starting points for a model where we account for the effects of gravity darkening on a rotationally flattened star. 

\subsubsection{Atmospheric parameters and helium abundance}\label{sect_1D_parameters}
The atmospheric parameters (\teff \ and \logg) and helium abundance by number, $y$ = $\cal{N}$(He)/[$\cal{N}$(H)+$\cal{N}$(He)], have been estimated by finding the best match between a set of observed H and He line profiles and a grid of rotationally-broadened, synthetic profiles computed using the non-LTE line-formation code DETAIL/SURFACE coupled with LTE Kurucz models (Butler \& Giddings \cite{butler85}; Giddings \cite{giddings81}).\footnote{The spectra were convolved using the ROTIN3 program in the SYNSPEC suite of routines (see {\tt http://nova.astro.umd.edu/Synspec43/synspec.html}).} Such a hybrid approach (LTE atmospheric models, but a full non-LTE treatment for the line formation) has been shown to be adequate for late O and early B-type stars on the main sequence (Nieva \& Przybilla \cite{nieva07}; Przybilla et al. \cite{przybilla11}). 

The analysis has been independently carried out on the SOPHIE and mean ELODIE spectrum. The results are listed separately in the following but, as only relatively small differences are found, the values eventually adopted will be the averaged ones. A SOPHIE spectrum of the narrow-lined O9\,V star 10 Lac was continuum normalised and broadened to \vsini \ = 390 \kms, as found for HD\,93521 based on Fourier techniques (Rauw et al. 2008). The continuum regions were subsequently defined and used to normalise the two spectra of HD\,93521.

A set of unblended helium features (6 \ion{He}{i} and 3 \ion{He}{ii} lines)\footnote{The relevant lines are He\,{\sc i} $\lambda\lambda$\,4026, 4388, 4471, 4713, 4922, 5016 and He\,{\sc ii}\,4542, 4686, 5412.} was chosen after an inspection of a spectral atlas and the 10 Lac spectrum which was taken as reference. Three narrow-lined stars were analysed to validate the procedures used to derive the atmospheric parameters and abundances: 10\,Lac (O9 V; \vsini \ = 25 \kms), $\tau$\,Sco (B0.2 V; \vsini \ = 8 \kms) and HD\,57682 (O9 IV; \vsini \ = 25 \kms). Such a differential approach is expected to minimise the systematic errors (see, e.g., Vrancken et al. \cite{vrancken}). These stars have their parameters and abundances derived using the same codes from classical techniques whereby \teff \ is based on the ionisation balance of various elements (namely He, C, N and Si), \logg \ on the fit of the wings of the Balmer lines and the abundances on curve-of-growth techniques (10\,Lac: this study; $\tau$\,Sco: Hubrig et al. \cite{hubrig08}; HD\,57682: Morel \cite{morel11}). The H and He features used may be contaminated by some weak metallic lines. The synthetic spectra were computed taking these lines into account and using the abundances derived from the curve-of-growth analysis. Some portions of the line profiles encompassing metal lines not modelled by DETAIL/SURFACE were masked out. For HD\,93521, the same metal abundances and microturbulence (\micro \ = 10 \kms) as for 10\,Lac were adopted. Model atmospheres with a helium abundance twice solar were used for HD\,93521 in accordance with the high abundance found (see below).

An iterative scheme was used to estimate \teff, \logg \ and $y$. The temperature is taken as the value providing the best fit to the \ion{He}{i} and \ion{He}{ii} features with the same weight given to these two ions. The gravity is determined by fitting the wings of the Balmer lines and $y$ by fitting the \ion{He}{i} features. This procedure was repeated until the gravity used to fit the helium lines was identical to the value yielded by the Balmer lines, and proved for the narrow-lined stars to provide the best match to the reference values obtained using classical methods. As already noted for HD\,93521 (Howarth \& Smith 2001), the fit was much better for H$\epsilon$ than for the other Balmer lines. However, the gravity estimated from H$\epsilon$ was indistinguishable from the mean value computed taking all lines into account. This analysis has also been repeated after convolving the observed spectra of the reference stars with a rotational broadening function with \vsini \ = 390 \kms. The results are shown in Table \ref{tab_abundances}. For the three standard stars, the mean differences between the reference values and the results obtained through the synthesis are: $\Delta$\teff \ = $+155 \pm 733$\,K, $\Delta$\logg \ = $-0.09 \pm 0.07$\,dex and $\Delta y$ = $+0.005 \pm 0.016$. We finally obtain the following parameters for HD\,93521 assuming a spherical star: \teff \ = $30\,900 \pm 700$\, K, \logg \ = $3.67 \pm 0.12$\,dex and $y = 0.18 \pm 0.02$ (statistical errors estimated based on the differences for the three standard stars between the parameters found and the reference ones, as well as the values derived for HD\,93521 using the two spectra). These results confirm the helium enrichment found by Lennon et al. (1991; $y = 0.20 \pm 0.05$) and Howarth \& Smith (2001; $y = 0.18 \pm 0.03$). A comparison for HD\,93521 between the observed and best-fitting He synthetic line profiles is shown in Fig.\ref{fig_he}. No significant differences are found between the temperature and helium abundance yielded by the He\,{\sc i} singlet and triplet transitions (see Najarro et al. \cite{najarro}).

\subsubsection{Metal abundances}\label{sect_1D_abundances}
This is the first time that the metal abundances are derived for this star. Determining its chemical composition could help establishing its birth place, as significantly subsolar abundances are expected if it was formed in situ far from the Galactic plane. 

\begin{figure}
\hspace*{0.2cm}
\includegraphics[width=90mm]{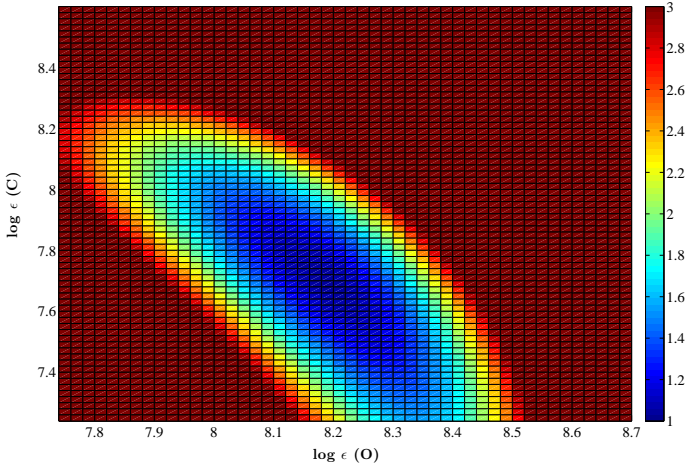}
\caption{Variation of the fit quality in the region 4060-4082 \AA \ for different combinations of the C and O abundances (colour coded as a function of the reduced $\chi^2$) in the case of the SOPHIE spectrum of HD\,93521. The best fit is found for $\log \epsilon$(C) = 7.70 dex and $\log \epsilon$(O) = 8.20 dex.}
\label{fig_C_O}  
\end{figure}

The high rotational velocity of HD\,93521 makes a selection of useful metallic features excessively difficult. As done for the hydrogen and helium lines, a number of spectral domains were carefully chosen and selected on the basis that: (1) at most two species should significantly contribute to the blend; (2) all the lines with a significant strength should be theoretically modelled; (3) all the elemental abundances derived for the three reference stars (either using the observed spectrum or the one broadened to the \vsini \ of HD\,93521) should reasonably match the reference values derived from the curve-of-growth analysis (the absolute difference is on average 0.09 dex and at most 0.23 dex). Only three spectral regions fulfilled these criteria: 4060-4082 \AA \ (main contributors \ion{C}{iii} and \ion{O}{ii}), 4691-4709 \AA \ (main contributor \ion{O}{ii}) and 4995-5011 \AA \ (main contributor \ion{N}{ii}). Unfortunately, the abundances of the other metals (e.g., Mg, Si) could not be reliably investigated. The parameters in Table \ref{tab_abundances} have been used, such that any errors in \teff \ and \logg \ are propagated to the derived abundances. The results are shown in Table \ref{tab_abundances}, while the observed and synthetic line profiles are shown in Fig.\ref{fig_fits}. Figure \ref{fig_C_O} shows an example of the variation of the fit quality in the region 4060-4082 \AA \ for different combinations of the C and O abundances. Taking into account the differences for the three reference stars between the abundances found and the reference ones, the different values obtained for oxygen from the two different spectral domains (there is a good agreement with a difference of at most 0.16 dex), and finally the values derived for HD\,93521 using the two spectra, the statistical errors on both the abundances and abundance ratios can be set at the $\approx$0.15 dex level.

Compared to the values derived for the three reference stars based on the spectra broadened with \vsini \ = 390 \kms, the abundances of HD\,93521 are on average lower by 0.69 dex for C, higher by 0.26 dex for N and lower by 0.27 dex for O. A nitrogen enrichment at the surface is expected given the high helium abundance and is indeed observed: the [N/C] and [N/O] abundance ratios are much higher than in 10\,Lac and HD\,57682, and also higher than in $\tau$\,Sco which is the prototype of the class of slowly-rotating, main sequence B stars (unexpectedly) displaying a nitrogen excess (e.g., Morel et al. \cite{morel08}). The quantitative behaviour of the [N/C] and [N/O] abundance ratios is within the errors what is expected for CNO-cycled material dredged up to the surface (Przybilla et al. \cite{przybilla10}
). Determinations of the CNO abundances for such fast rotators are very rare in the literature: Villamariz et al. (\cite{villamariz02}) found [N/C]$>$+0.78 and [N/O]$>$+0.12 dex for HD\,191423 (O9 III:n; \vsini \ = 450 \kms), while Villamariz \& Herrero (\cite{villamariz05}) found [N/C]$\approx$+0.48 and [N/O]$\approx$--0.35 dex for $\zeta$\,Oph (O9.5 Vnn; \vsini \ = 400 \kms).

\begin{table}[h!]
\centering
\caption{Sensitivity of the results to differences in continuum placement and choice of microturbulence. For the results with a different continuum location, the abundances have been estimated using the revised parameters.}
\begin{tabular}{l|ccc}
\hline\hline
                            & \multicolumn{2}{c}{Continuum shifted} & Microturbulence changed \\
                            & \multicolumn{2}{c}{upwards by}        & from 10 to 5 \kms\\
                            & 0.5\%  & 1\%    & \\\hline
$\Delta$ \teff \ [K]        & +490   & +1040  & ...\\
$\Delta$ \logg              & +0.10  & +0.24  & ...\\ 
$\Delta$ $y$                & +0.007 & +0.015 & ...\\ 
$\Delta$ $\log \epsilon$(C) & +0.10  & +0.14  & --0.03\\
$\Delta$ $\log \epsilon$(N) & +0.14  & +0.41  &  +0.05\\
$\Delta$ $\log \epsilon$(O) & +0.13  & +0.36  &  +0.12\\
$\Delta$ [N/C]              & +0.04  & +0.27  &  +0.08\\
$\Delta$ [N/O]              & +0.01  & +0.05  & --0.07\\\hline
\end{tabular}
\label{tab_sensitivity}  
\end{table}

As the abundances of the three metals investigated are affected at different extent by nuclear CNO-processing, it is difficult to estimate the metallicity of HD\,93521. However, oxygen is expected to be only slightly depleted at these levels of nitrogen enrichment and may be taken as a proxy for the global metal content. Assuming a depletion of oxygen of $\approx 0.1$\,dex (e.g., Heger \& Langer \cite{heger00}), there is thus some evidence for a lack of metals in HD\,93521 compared to solar at the 0.2\,dex level. It remains to be seen, however, if these low abundances are not an artifact of the data treatment: given the lack of true continuum regions, there may be a tendency to systematically place the continuum too low and therefore to underestimate the strength of the metal lines. To investigate this issue, we have redetermined the abundances of HD\,93521 after uniformly shifting the continuum level upwards by 0.5\%. This leads to an upward revision of \teff \ and \logg, as well as abundances that are typically 0.15\,dex higher (this translates to $\approx 0.30$\,dex for a shift of 1\%; Table \ref{tab_sensitivity}). We therefore conclude that the uncertainties in the metal abundances (especially owing to the difficulties in the continuum placement) prevent a clear conclusion to be drawn solely based on chemical arguments regarding the birthplace of HD\,93521 (either in or far from the Galactic plane). This is especially true when one considers that, although the differences are small, the assumption of a gravity darkened rotationally flattened star leads to systematically higher abundances (see Sect.\,\ref{sect_2D}). It should also be kept in mind that the microturbulent velocity is not constrained and has been fixed to $\xi$ = 10 \kms; adopting a lower value would result in generally slightly higher abundances (see  Table \ref{tab_sensitivity}). Nevertheless, our conclusions regarding the He and N-rich status of HD\,93521 appear robust against differences in continuum placement and choice of microturbulence (Table \ref{tab_sensitivity}).

\subsection{Assuming a gravity darkened rotationally flattened star\label{sect_2D}}
To refine the metal abundances, we have accounted for gravity darkening and rotational flattening using an  approach similar to the one of Howarth \& Smith (\cite{HS}). 

The stellar surface is assumed to be an equipotential surface of the Roche potential model. The latter can be written in spherical coordinates centred on the centre of a star of mass $m_*$.
\begin{equation}
\Omega  =  -\frac{G\,m_*}{r(\theta)} - \frac{\omega^2}{2}\times r^{2}\sin^{2}{\theta},
\end{equation}				
\noindent where $r$ is the distance from the centre of the star, $\theta$ is the colatitude, $\varphi$ is the longitude and $\omega$ the (constant) angular rotation velocity.

The local gravity is then equal to the gradient of the Roche potential. We have also take into account the inner radiation pressure following the approach of Howarth (\cite{Howarth97}). 
\begin{equation}
\vec{g} = (1 - \Gamma)\,\vec{\nabla} \Omega,
\end{equation}
\noindent where $\Gamma = \frac{\sigma_{Th}}{c}\sigma (T_{\rm eff}^p)^4\frac{1}{\left\Vert \vec{g}_{pole} \right\Vert}$,
 $\sigma_{Th} = 0.036$\,m$^2$\,kg$^{-1}$ is the Thomson scattering cross section, and $\sigma$ the Stefan-Boltzmann constant.
	
The temperature is computed using the von Zeipel (\cite{vonZeipel}) theorem
\begin{equation}
T_{\rm eff}^{local}=T_{\rm eff}^{p}\left(\frac{\left\Vert\nabla\Omega_{local}\right\Vert}{\left\Vert\nabla\Omega_{pole}\right\Vert }\right)^{0.25\,\alpha}
\end{equation}
\noindent where $\alpha$ is traditionally taken equal to 1 for massive stars. Recent interferometric observations of rapidly rotating B stars (e.g.\ Kraus et al.\ \cite{Kraus}), as well as theoretical work (Espinosa Lara \& Rieutord \cite{ELR}) suggest however that this might not be totally appropriate. Observations are better fitted by a value of $\alpha \approx 0.75$, whilst the theoretical results suggest a more complex law with a dependence of the best-fit $\alpha$ parameter on the actual flattening of the star. In the following, we have adopted $\alpha = 0.75$. This value of the gravity darkening exponent reduces the temperature contrast between the pole and the equator compared to the traditional von Zeipel law.

The stellar surface is discretized into $240 \times 60$ (longitude $\times$ colatitude) constantly spaced points. In each surface point, the local temperature and gravity are known, thus, we can compute a local contribution to the spectrum by linear interpolation between four spectra of a grid of synthetic spectra computed with the non-LTE line-formation code DETAIL/SURFACE coupled with the LTE Kurucz models. The spectra of the grid are spaced by 1000\,K in $T_{\rm eff}$ and 0.1\,dex in $\log{g}$. The appropriate Doppler shift is then applied to the local spectra accounting for the rotational velocity and these spectra are also multiplied by the area of the surface element projected along the line of sight towards the observer. The last corrective factor applied to the local contribution to the spectrum is the limb-darkening. The  limb-darkening coefficient is based on the tabulation of Al-Naimiy (\cite{Al-Naimiy}) for a linear limb-darkening law\footnote{We have compared the synthetic spectra with those simulated using more recent tabulations of the limb-darkening coefficients of Claret \& Bloemen (\cite{CB}). The differences between the normalized synthetic spectra are of the order $3 \times 10^{-4}$ continuum units, i.e.\ negligible in view of other uncertainties.}. The total spectrum is then computed by summing the incremental contributions of each surface point. We assume that there is no cross-talk between the different surface elements.

This model was then used to generate a grid of synthetic spectra for different metal abundances. To limit the number of free parameters, several parameters were frozen. This is the case of the stellar mass (15\,$M_{\odot}$), the effective temperature at the poles ($T_{\rm eff}^p = 34\,737$\,K), the polar radius (6.49\,$R_{\odot}$), the projected equatorial rotational velocity ($v\,\sin{i} = 390$\,km\,s$^{-1}$), the inclination of the rotation axis ($i = 90^{\circ}$) and the von Zeipel coefficient ($\alpha = 0.75$). The polar temperature and radius are chosen in such a way that the average effective temperature and \logg \ of the star as seen by the observer correspond to 30\,918\,K and $3.67$ respectively, i.e.\ agree with the numbers inferred in Sect.\,\ref{sect_1D_parameters}. This grid of models was then used in the same manner as the pure plane-parallel models of Sect.\,\ref{sect_1D_abundances} to adjust the three spectral regions (see Fig.\,\ref{fig_fits}) by varying the CNO abundances. The resulting CNO abundances are listed in the bottom part of Table\,\ref{tab_abundances}. As can be seen from this table, the differences between the abundances derived assuming either a spherical star or a rotationally flattened star are very small and the results of this section therefore fully confirm the conclusions of Sect.\,\ref{sect_1D_abundances}. This good agreement between the CNO abundances stems from the fact that in the rotationally flattened model we have chosen the average (area weighted) $T_{\rm eff}$ and $\log{g}$ to match the best-fit parameters of the pure plane-parallel model.

Finally, we have tested the sensitivity of the synthetic spectra computed from our rotationally flattened models to some of the frozen parameters. For this purpose, we have varied the mass to 8 and 20\,$M_{\odot}$, the inclination to $70^{\circ}$ and the von Zeipel coefficient to 0.50 and 1.0. In each case, the other parameters were adapted in such a way as to recover the mean surface temperature and gravity. The resulting differences in the synthetic spectra over the three spectral regions adopted for the CNO diagnostics, with respect to the best-fit model, were always found to be negligible.

\section{Discussion \label{discussion}}
\subsection{Could HD\,93521 be a hot subdwarf?}
Ebbets \& Savage (\cite{ES}) suggested that HD\,93521 could be a low-mass Population II star. Lennon et al.\ (\cite{Lennon}) argued that this is unlikely because of the strength of the metal lines that would be at odds with a low metal abundance expected for a Population II star. However, subdwarf O (sdO) and subdwarf B (sdB) stars are evolved low-mass objects (post red-giant branch, post horizontal branch or post asymptotic giant branch) which display a wide range of helium abundances and, in some cases, a strong enrichment of iron group elements (Heber et al.\ \cite{Heber}, Heber \cite{Heber2}). Thus the situation appears less clear cut than previously assumed. Therefore, it is worth to critically consider whether HD\,93521 could be a subdwarf OB star. 

The distinction between sdB and sdO stars stems from the presence of He\,{\sc ii} lines in the spectra of the latter and their absence in the former. Since HD\,93521 displays weak but definite He\,{\sc ii} lines in its spectrum, we have to compare its properties mostly with those of sdO stars, although the effective temperature of HD\,93521 that we infer from our model atmosphere fits would actually be more typical of an sdB star.  

Let us first consider the abundances. Most sdO stars display spectra that point at an He enrichment (up to a factor $10^4$ with respect to solar, e.g.\ Heber et al.\ \cite{Heber}) which is usually correlated with enhanced C and/or N abundances (Heber \cite{Heber2} and references therein). In the spectrum of HD\,93521, we find indeed a moderate enhancement of helium and nitrogen (Sect.\,\ref{optspec}), but at a less extreme level (for helium) than in typical sdOs. The abundance pattern of HD\,93521 could instead be the result of strong rotational mixing in a normal O-type star. Actually, the projected rotational velocity of 390\,km\,s$^{-1}$ is much larger than what is observed in hot subdwarfs. Indeed, with the exception of one object, single sdB stars are slow rotators with $v\,\sin{i} \leq 10$\,km\,s$^{-1}$ (Geier et al.\ \cite{Geier}). The exception is EC\,22081-1916 which displays $v\,sin{i} = 163$\,km\,s$^{-1}$ and could be the outcome of a merger event (see Geier et al.\ \cite{Geier}). Hot subdwarfs in very close binaries could be spun up by tidal forces, but the binary fraction among He-enriched sdOs is rather low (a few percent), much lower than in sdBs (Napiwotzki et al.\ \cite{Napiwotzki}). It seems very unlikely also that HD\,93521 could be in a close binary system: its large $v\,\sin{i}$ indicates that such a system would be seen under an inclination near  $i \approx 90^{\circ}$ and should hence display short-term radial velocity variations that would have been clearly detected in previous intensive spectroscopic monitoring campaigns (Rauw et al.\ \cite{Rauw}, see also Sect.\,\ref{runaway}). 

Hot ($T_{\rm eff} \geq 30000$\,K) subdwarfs of the sdB category are known to display short period (a few minutes) pressure mode non-radial pulsations. Much longer periods (45\,min to 2\,hrs) attributed to gravity modes were also observed in somewhat cooler sdBs ($T_{\rm eff} \leq 30000$\,K; Woudt et al.\ \cite{Woudt}, see also Fontaine et al.\ \cite{Fontaine} and references therein). There is only one known example of a pulsating sdO in the field (Woudt et al.\ \cite{Woudt}) which has very fast pulsations (periods of 1 -- 2 minutes). The non-radial pulsations detected in HD\,93521 have periods of 1.75 and 2.89\,hrs (Rauw et al.\ \cite{Rauw}) much longer than the p-mode oscillations seen in hot sdBs. Although they are comparable to the periods of slowly pulsating sdB stars, the effective temperature and gravity place HD\,93521 clearly outside the range of this category (see Fig.\,1 of Fontaine et al.\,\cite{Fontaine}). Our model atmosphere fits (Sect.\,\ref{optspec}) yield an averaged $\log{g} = 3.67$, much smaller than for typical sdOs ($\log{g}$ in the range 4.6 to 6.7, see Fig.\,2 of Heber et al.\ \cite{Heber}). 

HD\,93521 has an apparent magnitude of $V = 7.06$. Its {\it Hipparcos} parallax, $(0.85 \pm 0.49) \times 10^{-3}$\,arcsec (van Leeuwen \cite{vanLeeuwen}), indicates a distance that is significantly larger than for most other sdOs of comparable apparent magnitude (e.g.\ BD+75$^{\circ}$\,325, $V = 9.44$, $\Pi = (7.39 \pm 0.95) \times 10^{-3}$\,arcsec, sdO5; HD\,49798, $V = 8.18$, $\Pi = (1.20 \pm 0.50) \times 10^{-3}$\,arcsec, sdO5.5), although the error on the parallax of HD\,93521 and HD\,49798 are admittedly quite large. The luminosities of sdOs actually span two orders of magnitude (Heber \cite{Heber2}) and part of this wide range is due to their wide range of temperatures (36000 -- 78000\,K, Heber et al.\ \cite{Heber}). Our determination of the surface temperature would place HD\,93521 towards the border between sdO and sdB stars. This means that we have to compare the parallax of HD\,93521 with that of late-type and hence probably intrinsically fainter sdOs, thereby enhancing the disagreement.  

As an additional criterion for the sdO scenario we can consider the stellar wind features. The winds of sdOs are weak in comparison to those of normal O-stars (see e.g.\ Hamann et al.\ \cite{WRsdO}) and the associated UV spectral features are thus also generally weaker. {\it IUE} and {\it HST} spectra of HD\,93521 around the relevant wind lines are shown by Massa (\cite{Massa}), Howarth \& Reid (\cite{HR}) and Bjorkman et al.\ (\cite{Bjorkman}). There is a rather prominent P-Cygni feature in the C\,{\sc iv} $\lambda\lambda$\,1548, 1551 doublet which contrasts with the narrow, symmetric absorptions seen in the sdO star HD\,49798 (Hamann et al.\ \cite{WRsdO}). However, the profiles of the wind lines (including the C\,{\sc iv} doublet) of HD\,93521 are rather peculiar for O-type stars. These peculiarities have been interpreted as the signature of an axisymmetric wind structure with a density contrast of about 60 between the equatorial and polar wind (Bjorkman et al.\ \cite{Bjorkman}). The determination of the global mass-loss rate of HD\,93521 is thus not straightforward (spherically symmetric wind models obviously fail in this context). Using their axisymmetric model, Bjorkman et al.\ (\cite{Bjorkman}) quote the best-fit values of the product of the polar mass-loss rate times the ionization fraction for C\,{\sc iv}, Si\,{\sc iv} and N\,{\sc v}. These values range between $10^{-10.5}$ and $10^{-9.0}$\,$M_{\odot}$\,yr$^{-1}$, which, if the ionization fractions were close to unity, would be quite comparable to the mass-loss rates of the sdOs studied by Hamann et al.\ (\cite{WRsdO}). However, Bjorkman et al.\ (\cite{Bjorkman}) argue that the ionization fractions are significantly less than unity and that their result would be consistent with $\dot{M} = 10^{-7.2}$\,$M_{\odot}$\,yr$^{-1}$. On the other hand, the four sdOs studied by Hamann et al.\ (\cite{WRsdO}) are much hotter than HD\,93521 and a comparison with late sdO or sdB stars might be better. So far, observational evidence for winds in this category of subdwarfs is scarce and the few cases with a positive detection of wind emission in the H$\alpha$ line are consistent with mass loss rates of $10^{-11}$ to $10^{-10}$\,$M_{\odot}$\,yr$^{-1}$ (Unglaub \cite{Unglaub}). Therefore, our present knowledge of the wind of HD\,93521 does not allow us to draw a firm conclusion about the comparison of its mass-loss rate with that of hot subdwarfs.

Finally, we need to address the issue of the X-ray emission. Little is known about the X-ray emission of hot subdwarfs. An exception are two compact binary systems, HD\,49798 and BD$+37^{\circ}$\,442, which host a white dwarf or neutron star accreting material from the sdO component (Tiengo et al.\ \cite{Tiengo}, Mereghetti et al.\ \cite{Sandro, Sandro2}, La Palombara et al.\ \cite{Lapalombara}). The X-ray spectra of these objects are rather soft and are well fitted by a model consisting of a black-body component (kT $\approx 40$\,eV) along with a power-law ($\Gamma \approx 2$). Both objects show pulsations of the X-ray flux with periods of 13.2 and 19.2\,s. As we have shown in Sect.\,\ref{EPIC93521}, our X-ray data are not well described by this spectral model and do not display any signature of pulsations with periods between 1.0 and 1000\,s. This excludes the possibility that HD\,93521 could be an sdO with a compact companion.

During the eclipse of the white dwarf in HD\,49798, Mereghetti et al.\ (\cite{Sandro2}) detected an X-ray luminosity of $2 \times 10^{30}$\,erg\,s$^{-1}$, corresponding to $L_{\rm X}/L_{\rm bol} = 7 \times 10^{-8}$ which they attribute to the intrinsic X-ray emission of the sdO component. Whilst this $L_{\rm X}/L_{\rm bol}$ ratio is very similar to that of HD\,93521, and of normal O-type stars in general, overall the various arguments presented in this section are definitely at odds with an sdO or sdOB nature of HD\,93521.

\subsection{The runaway status \label{runaway}}
We have measured the radial velocity of HD\,93521 on our mean ELODIE spectrum. As this spectrum results from the combination of many individual spectra taken at various NRP phases, one expects it a priori to provide line profiles that should be rather free of the NRP signatures. We have only measured lines that are expected to be relatively free of strong blends and have a rather symmetric shape. Our line list is composed of H$\gamma$, H$\beta$, He\,{\sc i} $\lambda\lambda$\,4471, 4922 and He\,{\sc ii} $\lambda\lambda$\,4200, 4686, 5412. In this way, we infer a heliocentric radial velocity of $9.5 \pm 6.3$\,km\,s$^{-1}$, which, using the formalism of Moffat et al.\ (\cite{Moffat}) and Naz\'e (\cite{Yael}), translates into a peculiar radial velocity of $10.0 \pm 6.3$\,km\,s$^{-1}$, well below the conventional 30\,km\,s$^{-1}$ threshold (Cruz-Gonz\'alez et al.\ \cite{Cruz}) for runaway stars.

Using radial velocities compiled from the literature, Gies (\cite{Gies}) found a heliocentric radial velocity of $-13.9$\,km\,s$^{-1}$ corresponding to a peculiar radial velocity of $-12.5$\,km\,s$^{-1}$. Whilst this result is still below the threshold for a runaway status, it is actually quite different from our result. This prompted us to check the RVs on the average He\,{\sc i} line profiles in the data used for the analysis of the NRPs (Rauw et al.\ \cite{Rauw}). The latter yield heliocentric RVs of $-16.2$ and $-2.9$\,km\,s$^{-1}$ for the He\,{\sc i} $\lambda$\,5876 line respectively in February 2006 and April 2007, as well as $35.6$ and $3.8$\,km\,s$^{-1}$ for the He\,{\sc i} $\lambda$\,6678 line respectively in February 1997 and May 2005. Finally, measuring the same lines as for the mean ELODIE spectrum on our SOPHIE spectrum yields a heliocentric RV of $21.4 \pm 11.1$\,km\,s$^{-1}$. Whilst these differences could be due to orbital motion in a (most likely) long-period (years) binary system, an alternative explanation could be the NRPs. Rzaev \& Panchuk (\cite{RP}) studied the bisector radial velocities of a number of He\,{\sc i} lines in the spectrum of HD\,93521 over a time span of 2.7\,hrs. They found apparent variations of the radial velocities with amplitudes of tens of km\,s$^{-1}$. Therefore, the radial velocities of HD\,93521 could strongly depend on the specific phase of the NRPs or the sampling of the latter. This means that the true heliocentric RV is difficult to assess. Nonetheless, in view of the above results, it seems unlikely that the true peculiar velocity of HD\,93521 could be significantly larger than the 30\,km\,s$^{-1}$ threshold. 

The {\it Hipparcos} catalogue based on the revised reduction (van Leeuwen \cite{vanLeeuwen}) lists a parallax of $\Pi = (0.85 \pm 049) \times 10^{-3}$\,arcsec and proper motions of $\mu_{\alpha}\,\cos{\delta} = (0.32 \pm 0.40) \times 10^{-3}$ and $\mu_{\delta} = (2.44 \pm 0.37) \times 10^{-3}$\,arcsec\,yr$^{-1}$. We have used these numbers to estimate a peculiar tangential velocity of HD\,93521 of $v_t = 26.8 \pm 9.2$\,km\,s$^{-1}$. Adopting the criterion of Moffat et al.\ (\cite{Moffat}), this number is well below the threshold for considering HD\,93521 as a runaway star. 

Based on its kinematic properties, HD\,93521 can thus not unambiguously be classified as a runaway star. However, de Wit et al.\ (\cite{deWit}) argue that stars located more than 500\,pc above the Galactic plane must {\it de facto} be runaways, as there is currently no evidence for star formation in the halo. Indeed, in their study of a sample of 10 high Galactic latitude B-type stars with masses between 5 and 13\,$M_{\odot}$, Ramspeck et al.\ (\cite{Ramspeck}) did not find a conclusive candidate for a star formed in the halo. Although they were lacking kinematic data for some of their targets, they found that the evolutionary time scales were sufficiently long for the stars to have travelled to their current location if their ejection velocity reaches values up to 440\,km\,s$^{-1}$. 
\begin{figure}
\hspace*{0.2cm}
\includegraphics[width=85mm]{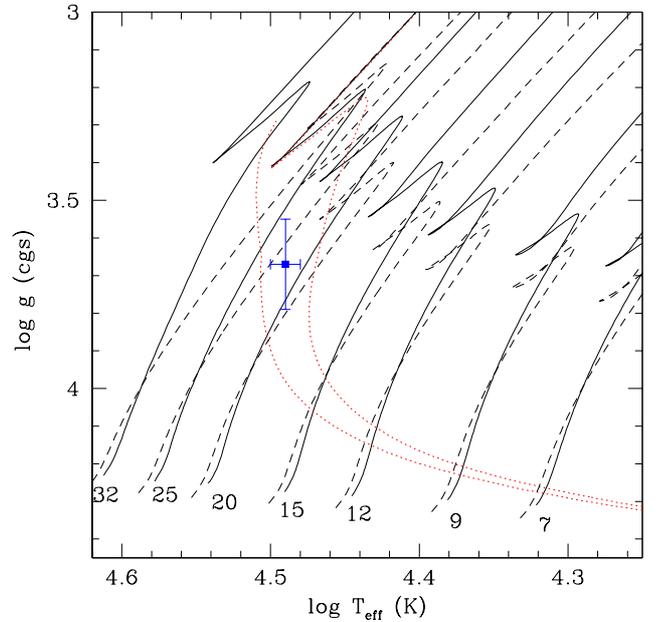}
\caption{Location of HD\,93521 in the $(\log{T_{\rm eff}}, \log{g})$ diagram (blue dot with error bars) as inferred from our model atmosphere fit (Sect.\,\ref{sect_1D_parameters}. The main-sequence parts of the Geneva evolutionary tracks from Ekstr\"om et al.\ (\cite{Ekstroem}) are shown by the solid lines for stars initially rotating at 40\% of the critical velocity. The initial mass of the star is given in $M_{\odot}$ by the label at the bottom of the track. The dashed lines yield the same tracks for zero rotational velocity. The isochrones for ages of 6.3 and 7.9\,Myr are shown by the dotted red lines.}
\label{isochrones}  
\end{figure}
To estimate the evolutionary age of HD\,93521, we have plotted the star in a $(\log{T_{\rm eff}}, \log{g})$ diagram against the evolutionary tracks and isochrones of Ekstr\"om et al.\ (\cite{Ekstroem}). The result is shown in Fig.\,\ref{isochrones}. HD\,93521 falls between the evolutionary tracks of initial mass 20 and 25\,$M_{\odot}$ and close to the 6\,Myr isochrone. Note that this estimate might actually be a lower limit of the true age, if the evolution of the star was affected by binary effects, such as mass transfer during a Roche lobe overflow phase or a merger event (following ejection through dynamical interactions in a dense cluster)\footnote{In the dynamical interactions scenario, the ejection of a close binary system is a rare event and should proceed at a comparatively low ejection velocity (Leonard \& Duncan \cite{LD}).} leading to a blue straggler star (de Mink et al.\ \cite{deMink}).

Dynamical interactions in dense clusters can produce very high maximum ejection velocities of up to 700\,km\,s$^{-1}$  for a 60\,$M_{\odot}$  O-star and even up to 1400\,km\,s$^{-1}$ for low-mass stars (Leonard \cite{Leonard}), whilst the supernova ejection scenario generally predicts maximum ejection velocities of $\leq 300$\,km\,s$^{-1}$ (Portegies-Zwart \cite{PZ}). However, Przybilla et al.\ (\cite{PNHB}) proposed that an ejection velocity of about 400\,km\,s$^{-1}$ could result from a supernova explosion in a tight Wolf-Rayet + B-star binary. From a sample of 96 likely OB runaway stars, Silva \& Napiwotzki (\cite{SN}) found evidence for a bi-modal distribution of the ejection velocities with two populations with velocities $\leq 300$\,km\,s$^{-1}$ and around 400 -- 500\,km\,s$^{-1}$ respectively. Therefore, assuming an ejection velocity of 400\,km\,s$^{-1}$, the minimum time requested to reach the star's present location, 1\,kpc above the Galactic plane, would be about 2.4\,Myr, which is well below the estimated evolutionary age. Therefore, in light of our current results, we cannot rule out the possibility that HD\,93521 could be a runaway star, probably observed near the apex of its orbit. 

\section{Summary and conclusions \label{conclusion}}
Our analysis of the X-ray and optical data of HD\,93521 provides new clues about the nature and origin of this star. We found no evidence of a stellar cluster or other tracers of star formation activity which makes it unlikely that the star has formed at its current position. The X-ray data did not reveal any evidence for a compact companion that could have been produced in a supernova explosion. The stellar parameters inferred from an analysis of the optical and X-ray spectrum are consistent with an `ordinary' Population I late O-type star and reject an interpretation of the star as a low-mass subdwarf. The He and N surface abundances are enhanced, as expected for a fast rotating evolved main-sequence O-type star (Meynet \& Maeder \cite{MM}). The X-ray spectra can be represented by a two-temperature optically thin thermal plasma model with a dominant component with kT $\approx 0.3$\,keV and a rather hot secondary plasma component with kT $\approx 3$\,keV. At first sight, the subsolar metallicity inferred from fitting the X-ray spectrum could be somewhat of a surprise. However, it is not uncommon to derive subsolar metallicities from the fit of CCD X-ray spectra of O-type stars. A prominent example of this situation is the population of OB stars in M\,17 (Broos et al.\ \cite{Broos}) which has a metallicity of $\approx 0.3\,$Z$_{\odot}$. 

Whilst the runaway status of the star remains ambiguous, the most likely explanation seems nevertheless that the star has not formed in the Galactic halo, but was rather ejected from the plane, either through dynamical interactions in a dense cluster or an unstable multiple system, or as a result of a supernova explosion in a binary system. The rapid rotation of the star, if not due to mass and angular momentum transfer in a binary system, could be the result of the dynamical interaction, maybe associated with a merger event (e.g.\ de Mink et al.\ \cite{deMink}) that would produce a blue straggler and could alleviate the flight time constraint. One open issue that remains with this scenario is how the star maintained its extreme rotational velocity well after the event that led to its ejection from the plane. A better knowledge of the proper motion and distance of the star, as will be provided by ESA's forthcoming {\it GAIA} mission will certainly be extremely beneficial to complete our understanding of this intriguing star. 

\acknowledgement{We are most grateful to our amateur colleagues Gaston Dessy (Biesme) and Jo\"el Bavais (Ath) who provided us with imaging observations of the field of HD\,93521. We further wish to thank Drs.\ Val\'erie Van Grootel and Sandro Mereghetti for enlightening discussions respectively about the general properties of sdOs and their X-ray emission in particular. We thank an anonymous referee for his/her constructive report. T.M.\ acknowledges financial support from Belspo for contract PRODEX-GAIA DPAC. G.R.\ and M.P.\ acknowledge support through the XMM/INTEGRAL PRODEX contract as well as by the Communaut\'e Fran\c caise de Belgique - Action de recherche concert\'ee - Acad\'emie Wallonie - Europe. We would like to thank K.\ Butler for making the NLTE line-formation codes DETAIL/SURFACE available to us.}

\end{document}